\documentclass[]{spie}  

\usepackage{amsmath,amsfonts,amssymb}
\usepackage{graphicx}
\usepackage[colorlinks=true, allcolors=blue]{hyperref}
\usepackage{amssymb}
\usepackage{geometry}
\usepackage[misc]{ifsym}
\usepackage{bbding}
\usepackage{amsmath}
\usepackage{graphicx}
\usepackage{multirow}
\usepackage{array}

\usepackage{float}
\usepackage{multicol}

\title{Synthetic Velocity Mapping Cardiac MRI Coupled with Automated Left Ventricle Segmentation}

\author[a,*]{Xiaodan Xing}
\author[a, b]{Yinzhe Wu}
\author[a, c]{David Firmin}
\author[a, c]{Peter Gatehouse}
\author[a,*]{Guang Yang}
\affil[a]{National Heart and Lung Institute, Imperial College London, London, UK}
\affil[b]{Department of Bioengineering, Imperial College London, London, UK}
\affil[c]{Cardiovascular Research Centre, Royal Brompton Hospital, London, UK}

\authorinfo{*Send correspondence to X. Xing (xxing@imperial.ac.uk) and G. Yang (g.yang@imperial.ac.uk).
}

\begin{document} 
\maketitle
\begin{abstract}
Temporal patterns of cardiac motion provide important information for cardiac disease diagnosis. This pattern could be obtained by three-directional CINE multi-slice left ventricular myocardial velocity mapping (3Dir MVM), which is a cardiac MR technique providing magnitude and phase information of the myocardial motion simultaneously. However, long acquisition time limits the usage of this technique by causing breathing artifacts, while shortening the time causes low temporal resolution and may provide an inaccurate assessment of cardiac motion. In this study, we proposed a frame synthesis algorithm to increase the temporal resolution of 3Dir MVM data. Our algorithm is featured by 1) three attention-based encoders which accept magnitude images, phase images, and myocardium segmentation masks respectively as inputs; 2) three decoders that output the interpolated frames and corresponding myocardium segmentation results; and 3) loss functions highlighting myocardium pixels. Our algorithm can not only increase the temporal resolution 3Dir MVMs, but can also generates the myocardium segmentation results at the same time.

\end{abstract}


\section{Description of purpose}
\begin{figure}
    \centering
    \includegraphics[width=16 cm]{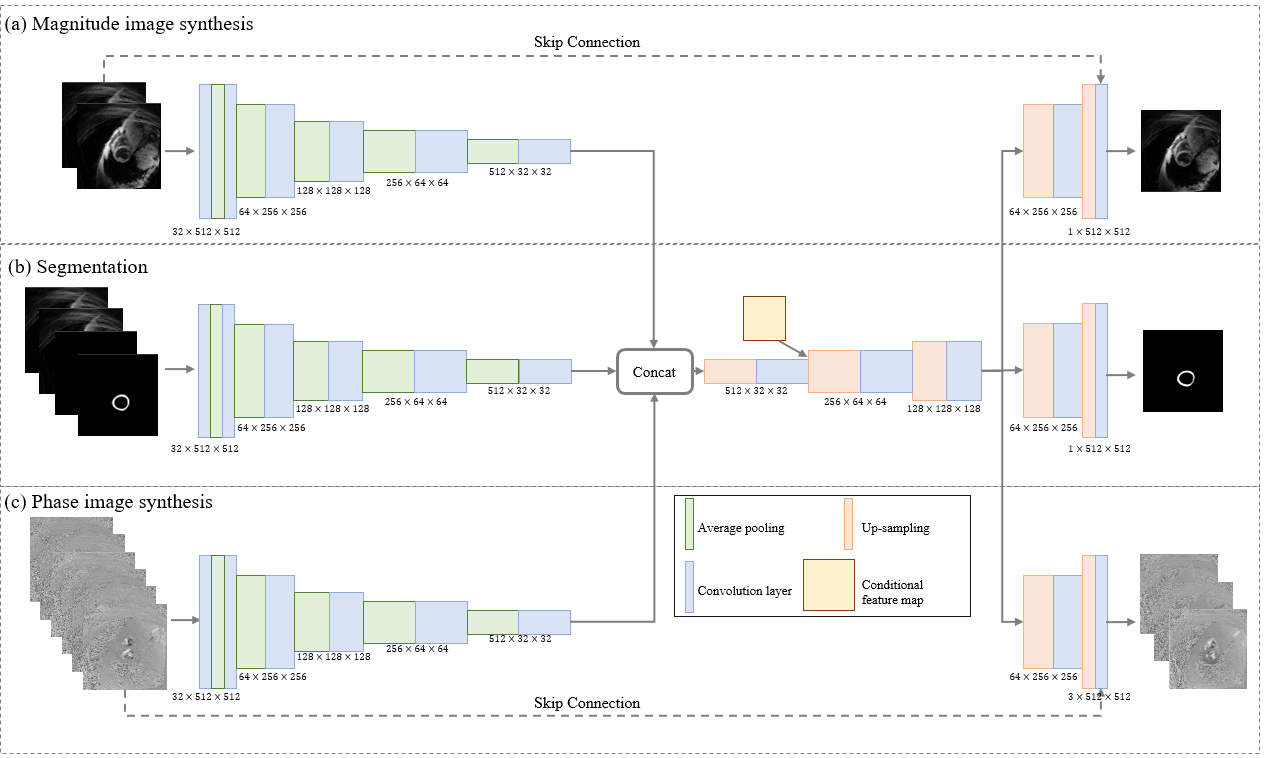}
    \caption{Multi-task attention UNet for different tasks. The resolution specified under each block is the resolution of the output feature map.}
    \label{fig:1}
\end{figure}

Three-directional cine multi-slice left ventricular myocardial velocity mapping (3Dir MVM) is a unique CMR technique that provides both spatial and temporal patterns of cardiac motion by acquiring magnitude images and velocity-encoded phase images correspondingly. Magnitude images, as shown in Figure \ref{fig:1}(a), provide the spatial location and shape information of left ventricles. Velocity encoded phase images in Figure \ref{fig:1}
(c), encodes the velocities of cardiac movement in three orthogonal directions \cite{lotz2002cardiovascular}. However, the application of this technique is limited due to the long acquisition time. Reducing the acquisition time while maintaining the temporal resolution becomes an interesting problem. To increase the temporal resolution of velocity mappings with a short acquisition time, generating the intermediate velocity mapping frames from existing frames is a potential a solution.

In this study, we propose a frame synthesis algorithm to increase the temporal resolution of 3Dir MVM data. Our aim is to synthesize unknown target images from their existing neighbors using a deep neural network. By improving the temporal resolution of 3Dir MVM data, we could reduce the acquisition time while not corrupting temporal patterns of the cardiac motion.

\noindent\textbf{Problem description:} We choose $t=\tau$ as a a start time point and $t=\tau + 4$ as an end time point. The goal of our project is to synthesize images at $t=\tau+k$, $k\in 	\left\{ 1,2,3 \right\}$. Since cardiac cycles for all healthy controls are similar, the relative time position of each image can provide decisive information of the shape of myocardium and surrounding tissues. Thus, our algorithm aims to achieve a conditional synthesis problem with $\tau$ and $k$ as conditions.  

\section{Method}
We introduce below the proposed multi-head multi-tail architecture for solving the synthesis problem. We develop an attention-based UNet structure \cite{oktay2018attention} and then split it into three threads with different encoders and decoders. Three encoders receive different inputs and the feature maps encoded are concatenated with a condition map. This condition map is a two-channel feature map that contains the normalized time point information of input images and the target image, which is aforementioned $\tau$ and $k$. To speed up the loss convergence, we incorporate skip connections at image synthesis stages.

\noindent\textbf{Data acquisition} We trained our algorithm on our in-house dataset \cite{simpson2013efficient}, which was obtained from Royal Brompton Hospital and contained 30 healthy subjects. The myocardium of all frames were manually segmented. Temporally, 50 frames were reconstructed per cardiac cycle. For each frame, one 2D magnitude image and three 2D phase images encoding velocities at three orthogonal directions were obtained. The resolution of each frame was $1.7\times 1.7 mm^2$ and was then re-sampled into $0.85\times 0.85 mm^2$ with a matrix size of $512\times 512$. We will address this series of frames as \emph{a series} in the following statement. Since 3Dir MVM had multiple slices, each subject had 3-5 slices acquired from base to apex of the left ventricle, which indicated that there were 3-5 series for each subject. We randomly selected 20 subjects (90 series) for training, 5 subjects (28 series) for validating and 5 subjects (27 series) for an independent testing. 

\noindent\textbf{Network parameters} Our model was trained on $512\times 512$ frames, but the network could be adaptable to all input resolutions because of the pooling layers. The two channel condition map had a size of $2\times 32 \times 32$. The learning rate was 0.001, and batch size was set to 32. To avoid information leakage between frames, we replaced batch normalization by instance normalization. 

\noindent\textbf{Loss functions and evaluation metrics} For synthesis, we observed that vanilla distance error over all pixels, such as Mean Absolute Error (MAE), might bring network to a local minimum because most pixels in magnitude images and phase images are noise. In this case, we used a weighted MAE $l_{MAE}$ to highlight the pixels within the region of interest (ROI). We used two weight maps: $\omega_1$ is the de-noising weight map, which is computed by original magnitude images and only non-background (non-black) pixels are highlighted; $\omega_2$ is the myocardium weight map, which is obtained by highlighting all pixels inside the epicardiums of all frames with a dilation of 2 pixels.

For segmentation, since the shape variation mainly occurs at the boundary of myocardium, we used the summation of the Dice loss $l_{DC}$ and the boundary loss  $l_{BD}$ \cite{kervadec2019boundary} as the loss function. The overall loss function is shown below:
\begin{equation}
    l = w_{syn}(\omega_1 +  \omega_2)l_{MAE} + w_{seg}(l_{DC} + l_{BD}) 
\end{equation}\\
Here $w_{syn}$ and $w_{seg}$ represent the weight parameters for the synthesis and the segmentation respectively. To evaluate the performance of our synthesis model, we used three quantitive metrics: 1) MAE; 2) Peak signal-to-noise ratio (PSNR); 3) Structural similarity (SSIM).   
\section{Results}
\begin{figure}
    \centering
    \includegraphics[width=\textwidth]{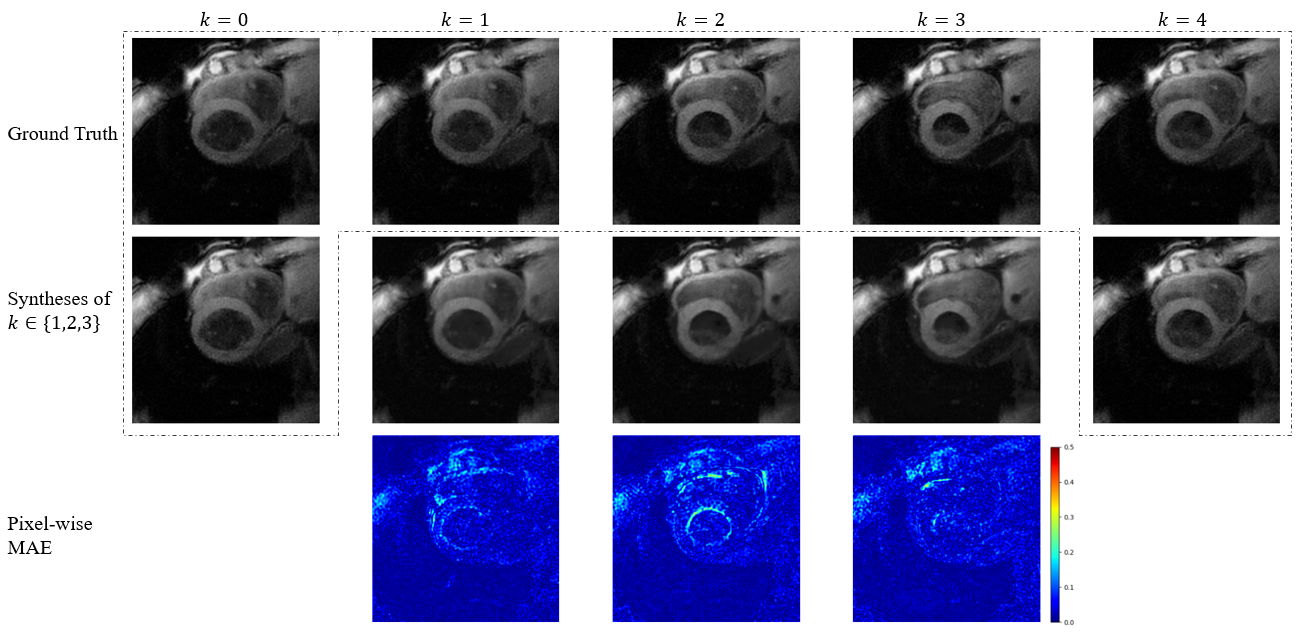}
    \caption{Example magnitude synthesis result. This example has a $\tau=0$. Magnitude images in dash line boxes are ground truth magnitude images, while others are synthesized results.}
    \label{fig:3}
\end{figure}
\begin{figure}
    \centering
    \includegraphics[width=15 cm]{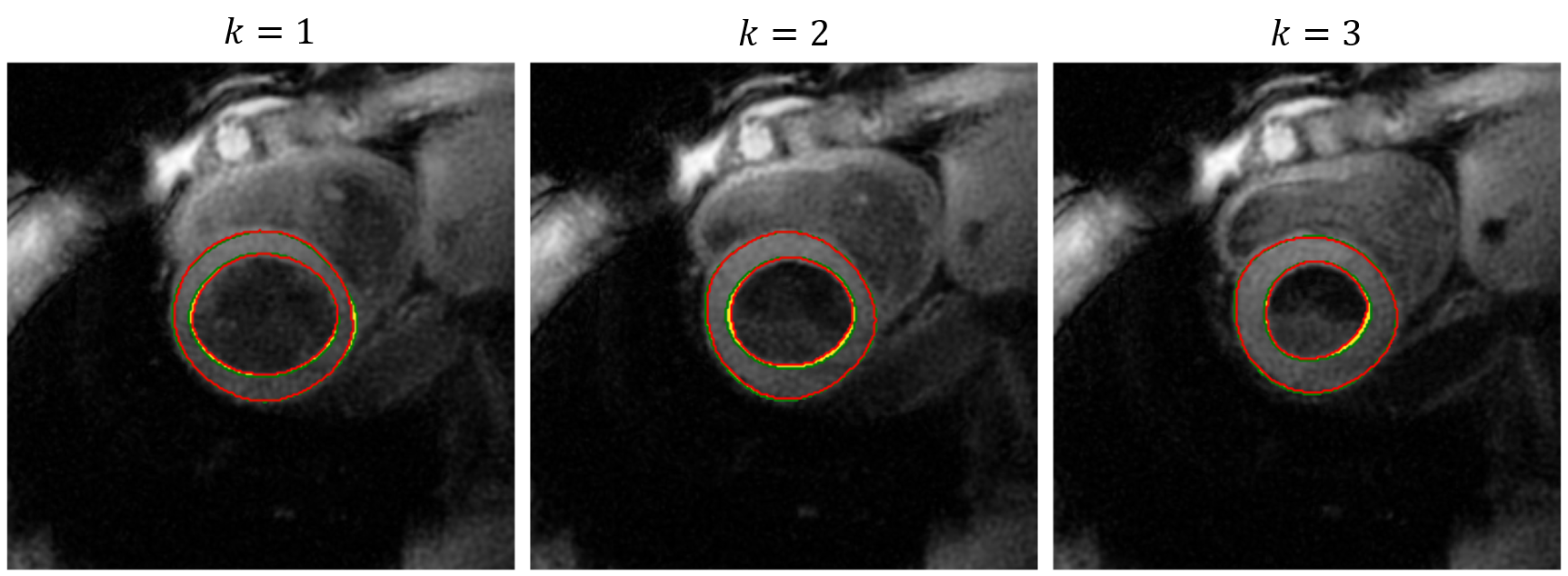}
    \caption{Segmentation results of one subject with $\tau=0$ and $k\in{1,2,3}$. Green contours are the ground truth segmentation result, while red contours are the predictions. The absolute difference between ground truth and prediction is highlighted with yellow.}
    \label{fig:4}
\end{figure}
\begin{figure}
    \centering
    \includegraphics[width=15 cm]{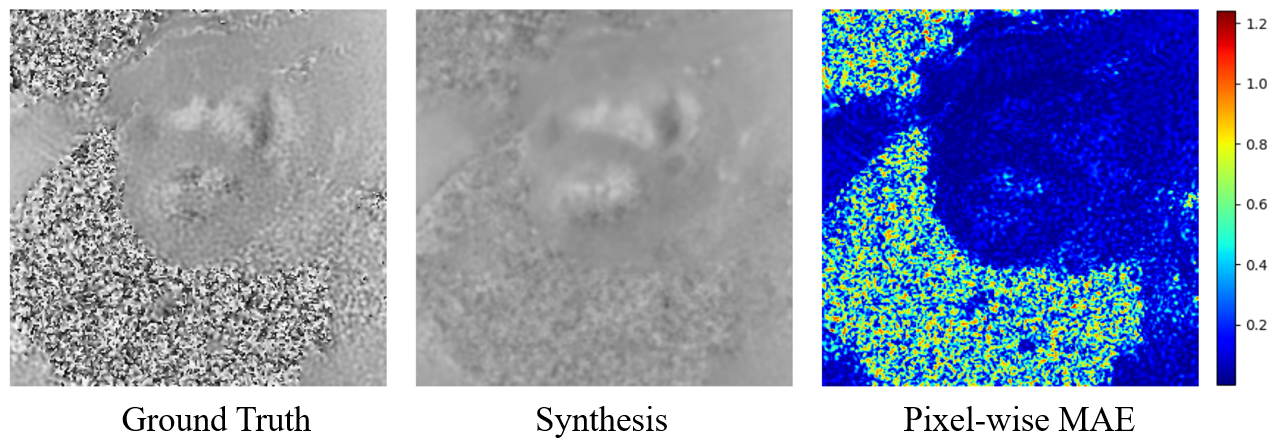}
    \caption{The phase synthesis result of one subject with $\tau=0$ and $k=1$.}
    \label{fig:5}
\end{figure}
Quantitative results of our proposed synthesis model are shown in Table \ref{tab:1}. We first compared our results with linear interpolation method and optical flow-based method, which was achieved by Horn-Schunck optical flow. Our algorithm could conditionally synthesize the intermediate 3Dir MVM frames with a promising accuracy on magnitude image. Example results of magnitude synthesis and corresponding segmentation and phase synthesis can be visualized in Figure \ref{fig:3}, \ref{fig:4} and \ref{fig:5}.

\begin{table}
\centering
\caption{Comparison methods.}\label{tab:1}
\begin{tabular}{m{2cm}<{\centering}m{2cm}<{\centering}m{2cm}<{\centering}m{2cm}<{\centering}m{2cm}<{\centering}m{2cm}<{\centering}m{2cm}<{\centering}}
\hline
Modality &	\multicolumn{3}{c}{Magnitude}&		\multicolumn{3}{c}{Phase}\\
\hline
Methods&	MAE&	PSNR&	SSIM&		MAE&	PSNR&	SSIM\\
\hline
Linear Interpolation&	0.037&		76.638&	0.841&		0.299&		58.696&	0.139\\
\hline
Horn-Schunck optical flow&	0.027&		79.246&	0.871&		0.363&		56.916&	0.158	\\
\hline
Ours&	\textbf{0.017}&		\textbf{83.505}&	\textbf{0.929}&		\textbf{0.166}&		\textbf{63.727}&	\textbf{0.225}	\\
\hline
\end{tabular}
\end{table}

We then validated the performance of the proposed modules in our model via ablation studies. For ablation studies, we added two metrics to evaluate the performance of predicted segmentation. To evaluate our segmentation, Dice scores were used, which can also be considered as an additional validation for our synthesis tasks. The velocity coefficient is calculated based on the velocity curves. Using segmentation masks and phase images, we could calculate the longitudinal, radial and circumferential velocity of the myocardium with a self-developed Python toolkit. The velocity coefficient is calculated by the averaged Pearson's correlation between predicted and true velocities. 
\begin{table}
\centering
\caption{Ablation studies.}\label{tab:2}
\begin{tabular}{m{1.7cm}<{\centering}m{1.7cm}<{\centering}m{1.5cm}<{\centering}m{1.5cm}<{\centering}m{1.7cm}<{\centering}m{1.2cm}<{\centering}m{1.2cm}<{\centering}m{2cm}<{\centering}m{2cm}<{\centering}}
\hline
Independent encoder& Independent decoder& Shared hidden layer& Weighted loss function&	Magnitude PNSR&	Phase PSNR&	Dice coefficient&	Velocity coefficient\\
\hline
\checkmark & \checkmark& 
$\times$&  \checkmark&		73.368&	59.982&	0.940&	0.841	\\

\hline
$\times$ & $\times$ & 
\checkmark & \checkmark&		76.654&	61.746&	0.945&	0.868\\
\hline
\checkmark & \checkmark & 
\checkmark & $\times$ &		76.487&	59.272&	0.945&	0.832\\
\hline
\checkmark & \checkmark &
\checkmark & \checkmark&	\textbf{83.505}&	\textbf{63.727}&	\textbf{0.959}&	\textbf{0.889}\\
\hline
\end{tabular}
\end{table}


\section{Conclusions}
In this study, we have proposed a multi-task attention UNet structure to synthesis intermediate 3Dir MVM frames using their existing neighbors. The experimental results have shown that our algorithm has not only improved the temporal resolution of 3Dir MVM data by synthesizing both magnitude and phase images, but has also generated corresponding left ventricle segmentation automatically and accurately. Upon acceptance of this abstract, we will provide experimental results on more comparison methods, as well as more evaluation analysis results. 

It is of note that although our algorithm has achieved promising performance on magnitude synthesis and segmentation generation, its performance on the synthesis of phase images still needs to be improved. Phase images lack structural visual information, or edges, compared to magnitude images, whereas convolution operations focus on edges of objects. In the future, we will focus on improving the performance of phase image synthesis. Our future work will also include investigation of the application of our algorithm from a single modality synthesis into cross-modality syntheses, such as synthesizing phase images from magnitude images. \\

\bibliographystyle{spiebib}
\bibliography{bibtex.bib}
\end{document}